\documentclass[apj]{emulateapj}

\shorttitle{Temperatures of Active Region Outflows}
\shortauthors{Warren et al.}

\begin{document}

%% ------------------------------------------------------------------------------------------
%% --- TITLE PAGE ---------------------------------------------------------------------------
%% ------------------------------------------------------------------------------------------

\title{The Temperature Dependence of Solar Active Region Outflows}

\author{Harry P. Warren, Ignacio Ugarte-Urra\altaffilmark{1}, Peter
  R. Young\altaffilmark{1}, and Guillermo Stenborg\altaffilmark{2}}

\affil{Space Science Division, Naval Research Laboratory, Washington, DC 20375}

\altaffiltext{1}{also College of Science, George Mason University, 4400 University Drive,
  Fairfax, VA 22030}

\altaffiltext{2}{also Interferometrics, Inc., Herndon, VA, 20171}

%% ------------------------------------------------------------------------------------------
%% --- ABSTRACT -----------------------------------------------------------------------------
%% ------------------------------------------------------------------------------------------

\begin{abstract}
  Spectroscopic observations with the EUV Imaging Spectrometer (EIS) on \textit{Hinode}
  have revealed large areas of high speed outflows at the periphery of many solar active
  regions. These outflows are of interest because they may connect to the heliosphere and
  contribute to the solar wind. In this Letter we use slit rasters from EIS in combination
  with narrow band slot imaging to study the temperature dependence of an active region
  outflow and show that it is more complicated than previously thought. Outflows are
  observed primarily in emission lines from \ion{Fe}{11}--\ion{Fe}{15}.  Observations at
  lower temperatures (\ion{Si}{7}), in contrast, show bright fan-like structures that are
  dominated by downflows. The morphology of the outflows is also different than that of
  the fans. This suggests that the fan loops, which often show apparent outflows in
  imaging data, are contained on closed field lines and are not directly related to the
  active region outflows.
\end{abstract}


%% ------------------------------------------------------------------------------------------
%% --- BODY ---------------------------------------------------------------------------------
%% ------------------------------------------------------------------------------------------

  \section{Introduction}

  The detection of large areas of high speed outflows at the periphery of many solar
  active regions is one of the more intriguing discoveries of the EUV Imaging Spectrometer
  (EIS) on \textit{Hinode}
  \citep[e.g.,][]{doschek2007,doschek2008,harra2008,delzanna2008}. These outflows are
  characterized by bulk shifts in the line profile of up to 50\,km~s$^{-1}$ and
  enhancements in the blue wing of up to 200\,km~s$^{-1}$
  \citep{bryans2010,mcintosh2009}. These outflow regions are of interest because they may
  lie on open field lines that connect to the heliosphere and contribute to the solar wind
  \citep[e.g.,][]{schrijver2003}.

  One aspect of the active region outflows that has not been investigated in sufficient
  detail is their temperature dependence. Spectroscopic observations have generally
  identified outflows in relatively hot coronal lines (\ion{Fe}{12}--\ion{Fe}{15}).
  Several studies, however, have associated active region outflows with the apparent
  motions observed along cooler, fan-like structures using imaging instruments
  \citep[e.g.,][]{sakao2007,hara2008,mcintosh2009,he2010}.

  In this Letter we use a combination of slit rasters and narrow band slot images to study
  the temperature dependence of outflows in an active region. Using EIS slit rasters we
  find that at lower temperatures (\ion{Si}{7}) the emission from the fans is strongly
  red-shifted, suggesting that this downflowing material lies on closed field lines. This
  interpretation is consistent with larger field of view images which suggest that the fan
  loops connect to the quiet Sun or to other active regions. At intermediate temperatures
  (\ion{Fe}{10}) we see a mixture of outflows and downflows. Outflows are observed from
  \ion{Fe}{11}--\ion{Fe}{15}. The velocities presented here are based on new relative rest
  wavelengths derived from EIS observations from the quiet corona above the limb. EIS slot
  imaging, which provides observations of 1\,\AA\ spectral bands over a wider field of
  view, is consistent with these findings. EIS slot movies of this region in \ion{Fe}{12}
  show apparent outflows while slot movies in \ion{Si}{7} show apparent downflows.

\section{Observations}

  \begin{figure*}[t!]
  \centerline{\includegraphics[clip,scale=0.975]{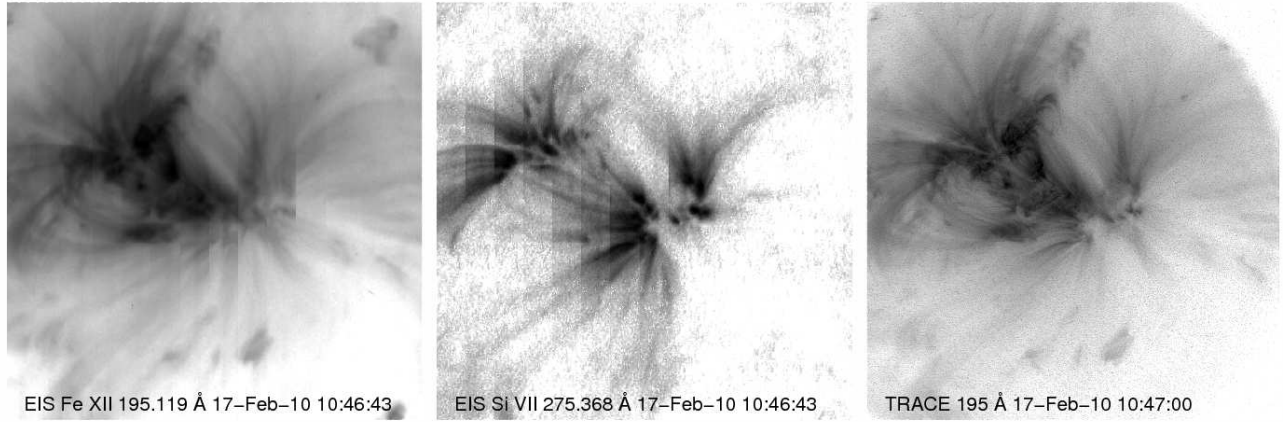}}
  \caption{(\textit{left panels}) EIS slot rasters in \ion{Fe}{12} 195.119\,\AA\ and
    \ion{Si}{7} 275.368\,\AA. The slot rasters are constructed by stepping the 40\arcsec\
    slot across the active region. (\textit{right panel}) An approximately co-temporal
    TRACE 195\,\AA\ image. The electronic version of the manuscript contains EIS and TRACE
    movies. The EIS slot movies show apparent outflows in \ion{Fe}{12} and apparent
    downflows in \ion{Si}{7}. The intensities are shown with a reversed color table where
    bright features are black. }
  \label{fig:movies}
  \end{figure*}

  \begin{figure}
  \centerline{\includegraphics[clip,scale=0.95]{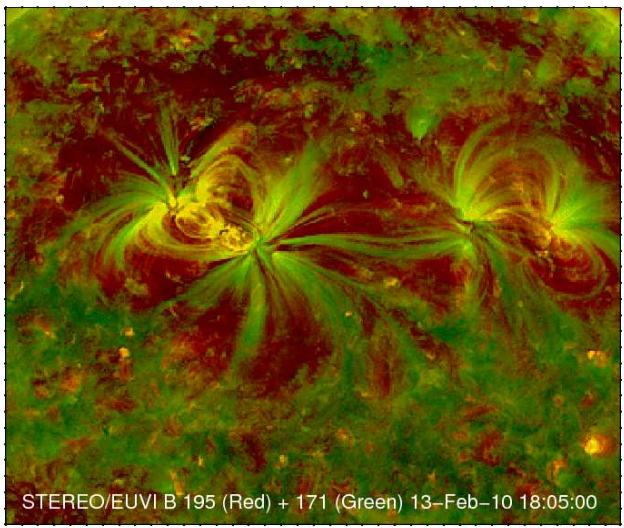}}
  \caption{A two-color, wavelet enhanced EUVI image of active region 11048. The 195\,\AA\
    is red and the 171\,\AA\ image is green. This image illustrates the various types of
    connections for the fan loops. Some fan loops connect to another active region, some
    loops connect to the quiet Sun, while others connect to the opposite polarity flux
    within the active region itself. The electronic version of the manuscript contains an
    animation of these data from 7--21 February 2010.}
  \label{fig:euvi}
  \end{figure}

  The EIS instrument on \textit{Hinode} is a high spatial and spectral resolution imaging
  spectrograph. EIS observes two wavelength ranges, 171--212\,\AA\ and 245--291\,\AA, with
  a spectral resolution of about 22\,m\AA\ and a spatial resolution of about 1\arcsec\ per
  pixel. There are 1\arcsec\ and 2\arcsec\ slits as well as 40\arcsec\ and 266\arcsec\
  slots available.  Solar images can be made using one of the slots or by stepping one of
  the slits over a region of the Sun. Telemetry constraints generally limit the spatial
  and spectral coverage of an observation. See \cite{culhane2007} and \cite{korendyke2006}
  for more details on the EIS instrument.

  For this analysis we consider two different EIS observing sequences that were run on
  NOAA active region 11048. One sequence involves constructing a raster by stepping the
  1\arcsec\ slit across a small region of the Sun ($180\arcsec\times512\arcsec$) taking a
  50\,s exposure at each position. To accelerate the observing a 3\arcsec\ step was taken
  between exposures.  Emission lines of interest were selected and fit with a single
  Gaussian. Measuring absolute velocities with EIS is difficult and we will discuss this
  in detail in the next section. This observing sequence was run 6 times during a two day
  period. Here we present results from the run that began at 04:25 UT and ended at 05:17
  UT on 17 February 2010.

  The second observing sequence involves constructing a series of relatively high cadence
  rasters using the 40\arcsec\ slot. The 40\arcsec\ slot provides a dispersed image
  similar to those taken with the SO82-A instrument on \textit{Skylab}
  \citep{tousey1977}. The relatively narrow 40\arcsec\ width of the EIS slot, however,
  yields images which integrate over only 1\,\AA\ of the solar spectrum. This limits the
  amount of overlap between images. To build up a larger field of view the slot is stepped
  across the Sun and, for this observing program, a 10\,s exposure is taken at each
  position. The observing program of interest here took a series of 14 exposures for each
  raster. The final field of view is about $448\arcsec\times480\arcsec$. The time between
  successive slot rasters is about 180\,s.  In this paper we present results from a run
  that began at 10:45 UT and ended at 15:30 UT on 17 February 2010.

  We also consider context observations from the \textit{Transition Region and Coronal
    Explorer} (\textit{TRACE}, see \citealt{handy1999}) and the EUV Imager (EUVI) on the
  \textit{Solar Terrestrial Relations Observatory} (\textit{STEREO}, see
  \citealt{howard2008}). Both \textit{TRACE} and EUVI are high resolution EUV multi-layer
  telescopes with four different channels. For both instruments there are three channels
  for imaging the corona: \ion{Fe}{9}/\ion{Fe}{10} 171\,\AA, \ion{Fe}{12} 195\,\AA, and
  \ion{Fe}{15} 284\,\AA. The \textit{TRACE} observations of interest here consisted
  primarily of full resolution 195\,\AA\ images. The median cadence was 47\,s, but there
  were several extended data gaps. These observations were taken from 10:00 UT to 16:00 UT
  on 17 February 2010. The EUVI images we have studied are from the behind spacecraft and
  show the evolution of the active region of interest from 7--21 February 2010. The
  separation angle between Earth and \textit{STEREO} B was about 71$^\circ$ on 17 February
  2010.

  \begin{figure*}[t!]
  \centerline{\includegraphics[clip,scale=1.0]{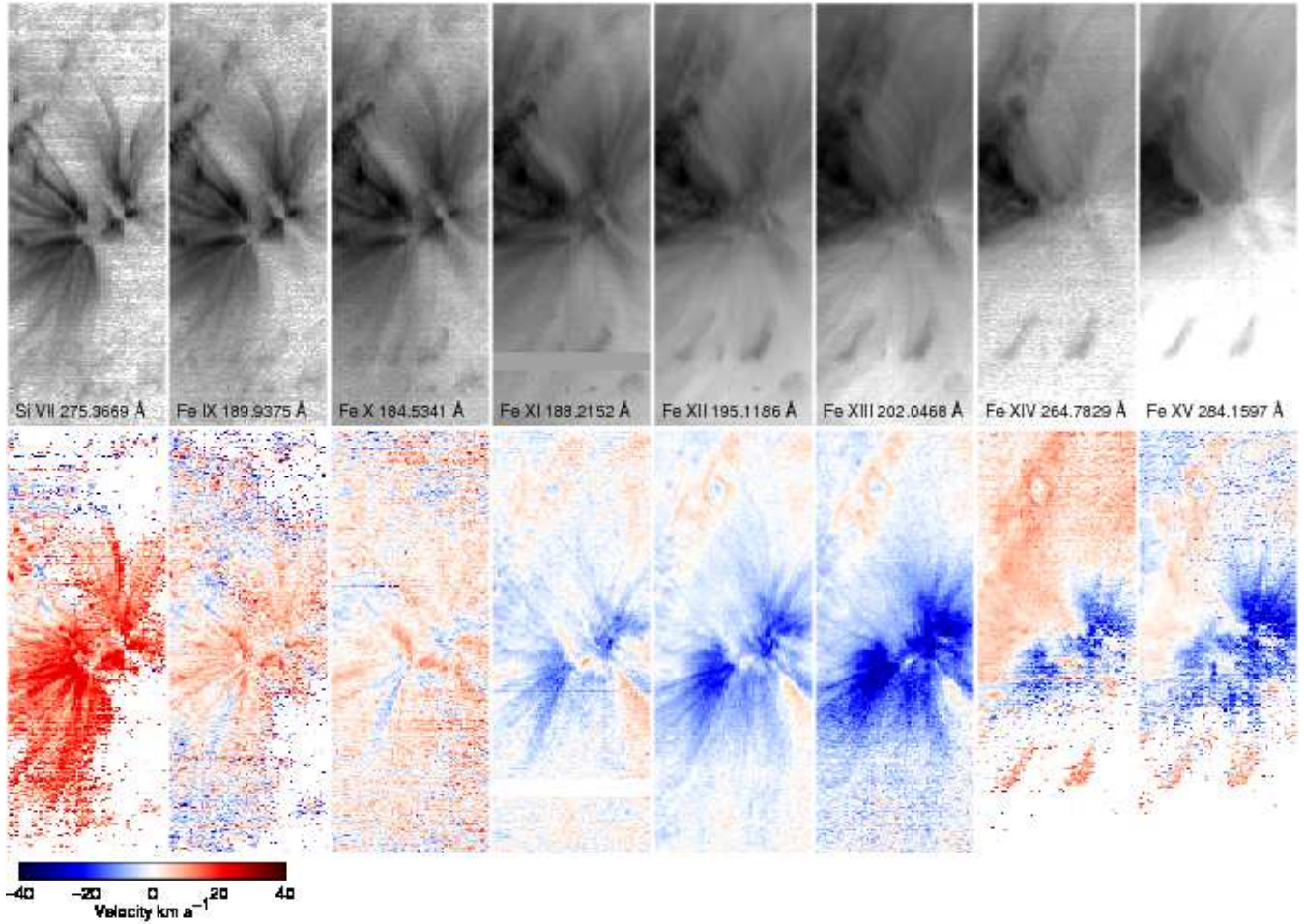}}
  \caption{EIS observations of active region outflows on 2010 February 17. The top panels
    show the intensities derived from Gaussian fits to the line profiles. The intensities
    are shown with a reversed color table. The bottom panels show the Doppler shifts
    relative to the quiet Sun in each line. Velocities for very weak line profiles have
    been ``grayed out''. The relative wavelengths are from the measurements presented in
    Table~\protect{\ref{table:lambda}}. The velocities show a progression from downflows at low
    temperatures to strong outflows at higher temperatures with a transition region around
    \ion{Fe}{10}. }
  \label{fig:flows}
  \end{figure*}

  \section{Results}

  EIS slot rasters from the \ion{Fe}{12} 195.119\,\AA\ and \ion{Si}{7} 275.368\,\AA\
  emission lines are shown in Figure~\ref{fig:movies}.  Movies of these EIS data are
  available in the electronic version of the manuscript. The movies show clear
  evidence for dynamical behavior in both the cool fan loops that are bright in
  \ion{Si}{7} and in the faint regions seen in \ion{Fe}{12}. The apparent motions seen in
  \ion{Si}{7} are suggestive of downflows. The apparent motions in \ion{Fe}{12}, in
  contrast, suggest highly episodic outflows.

  Also shown in Figure~\ref{fig:movies} is a nearly simultaneous \textit{TRACE} 195\,\AA\
  image from this region.  The \textit{TRACE} movie of these data, which is included in
  the electron version of the manuscript, is generally consistent with the EIS slot data
  and shows apparent outflows. We note that the while the fans are generally observed at
  lower temperatures, they can also appear in 195\,\AA\ images.  The 195\,\AA\ bandpass
  includes some cooler emission lines, such as \ion{Fe}{8} 194.663\,\AA, which can be
  comparable in intensity to the \ion{Fe}{12} 195.119\,\AA\ line in the fan loops
  \citep[e.g.,][]{landi2009,delzanna2003}. Thus the temperature of the emission imaged in
  the 195\,\AA\ channel is ambiguous. 

  The connectivity of the fan loops is unclear in the small field of view EIS and
  \textit{TRACE} images. In Figure~\ref{fig:euvi} we show a larger field of view of this
  active region extracted from a full Sun EUVI image. This image is a combination of
  wavelet enhanced 195\,\AA\ and 171\,\AA\ images. The multi-scale wavelet processing
  enhances the loops by removing the diffuse background and sharpening the remaining
  signal \citep{stenborg2008}. These images suggest that the fan loops lie on closed field
  lines and illustrate the various types of connections that these loops can have. The fan
  loops appear to connect to flux within the active region, other active regions, and the
  quiet Sun. A movie of these data are available in the electronic version of the
  manuscript.

  EIS intensity maps derived from the slit observations are shown in
  Figure~\ref{fig:flows}. As expected from the narrow bandpasses of the EIS slot images,
  the intensity maps show essentially the same morphology as the corresponding slot
  images. The additional lines available with the slit allow us to study the morphology of
  the flows as a function of temperature. The fan loops are dominated by cool emission and
  are bright in \ion{Si}{7}, \ion{Fe}{9}, and \ion{Fe}{10}. The morphology of the outflow
  region seen in the \ion{Fe}{12} 195.119\,\AA\ slot movies is echoed in all of the
  emission at intermediate temperatures (\ion{Fe}{11}--\ion{Fe}{13}). At the highest
  temperatures considered here (\ion{Fe}{14} and \ion{Fe}{15}) the emission is relatively
  weak in the outflow region, although generally above the level of the quiet Sun.

  Also shown in Figure~\ref{fig:flows} are the Doppler velocity maps derived from the slit
  data. The Doppler velocity maps in \ion{Si}{7} and \ion{Fe}{12} are consistent with the
  apparent motions seen in the movies, with strong downflows in \ion{Si}{7} and strong
  outflows in \ion{Fe}{12}. The Doppler maps also suggest that there is a transition from
  downflows to outflows around \ion{Fe}{10}. Our observations of strong downflows at low
  temperatures in the fans are consistent with previous measurements of downflows in
  active region fan loops in \ion{Ne}{8} by \cite{winebarger2002}.

% ll@{\hspace{1ex}$\pm$\hspace{1ex}}rrcll@{\hspace{1ex}$\pm$\hspace{1ex}}rr

\begin{deluxetable}{lcccrr}
\tablewidth{3.1in}
\tabletypesize{\scriptsize}
\tablecaption{Relative Wavelengths Observed With EIS\tablenotemark{a}}
\tablehead{
 \multicolumn{1}{c}{Line} &
 \multicolumn{1}{c}{$\lambda$} &
 \multicolumn{1}{c}{$\sigma_\lambda$} &
 \multicolumn{1}{c}{$\sigma_v$} &
 \multicolumn{1}{c}{$\delta_\lambda$} &
 \multicolumn{1}{c}{$\delta_v$} 
}
\startdata
      \ion{Fe}{8} 185.213 &   185.2107 &        0.5 &        0.9 &        2.3 &        3.7 \\
      \ion{Fe}{8} 186.601 &   186.6060 &        0.7 &        1.1 &       -5.0 &       -8.0 \\
      \ion{Fe}{8} 194.663 &   194.6523 &        0.7 &        1.1 &       10.7 &       16.5 \\
      \ion{Fe}{9} 189.941 &   189.9375 &        0.4 &        0.7 &        3.5 &        5.5 \\
      \ion{Fe}{9} 197.862 &   197.8570 &        0.3 &        0.5 &        5.0 &        7.5 \\
     \ion{Fe}{10} 184.536 &   184.5341 &        0.4 &        0.6 &        1.9 &        3.0 \\
     \ion{Fe}{11} 180.401 &   180.3990 &        0.3 &        0.5 &        2.0 &        3.4 \\
     \ion{Fe}{11} 188.216 &   188.2152 &        0.2 &        0.2 &        0.8 &        1.3 \\
     \ion{Fe}{11} 188.299 &   188.3004 &        0.2 &        0.2 &       -1.4 &       -2.2 \\
     \ion{Fe}{12} 192.394 &   192.3940 &        0.0 &        0.0 &        0.0 &        0.0 \\
     \ion{Fe}{12} 193.509 &   193.5090 &        0.1 &        0.2 &        0.0 &        0.1 \\
     \ion{Fe}{12} 195.119 &   195.1186 &        0.1 &        0.2 &        0.4 &        0.7 \\
     \ion{Fe}{13} 202.044 &   202.0468 &        0.2 &        0.2 &       -2.8 &       -4.1 \\
     \ion{Fe}{13} 203.826 &   203.8243 &        0.6 &        0.9 &        1.7 &        2.5 \\
      \ion{Si}{7} 275.368 &   275.3669 &        0.7 &        0.7 &        1.1 &        1.2 \\
     \ion{Si}{10} 253.791 &   253.7874 &        1.1 &        1.3 &        3.6 &        4.2 \\
     \ion{Si}{10} 258.375 &   258.3750 &        0.0 &        0.0 &        0.0 &        0.0 \\
     \ion{Si}{10} 261.058 &   261.0573 &        0.6 &        0.6 &        0.7 &        0.8 \\
     \ion{Si}{10} 271.990 &   271.9902 &        0.4 &        0.4 &       -0.2 &       -0.3 \\
     \ion{Si}{10} 277.265 &   277.2611 &        0.6 &        0.6 &        3.9 &        4.2 \\
     \ion{Fe}{14} 211.316 &   211.3192 &        0.6 &        0.9 &       -3.2 &       -4.6 \\
     \ion{Fe}{14} 264.787 &   264.7829 &        0.6 &        0.7 &        4.1 &        4.7 \\
     \ion{Fe}{14} 270.519 &   270.5223 &        0.4 &        0.5 &       -3.3 &       -3.7 \\
     \ion{Fe}{14} 274.203 &   274.2021 &        0.7 &        0.7 &        0.9 &        0.9 \\
     \ion{Fe}{15} 284.160 &   284.1597 &        1.0 &        1.0 &        0.3 &        0.3 \\
\enddata
\tablenotetext{a}{The literature wavelengths are from \protect{\cite{brown2008}}. The new
  wavelengths ($\lambda$) are in \AA. The 1-$\sigma$ variances in the measured wavelength
  is given in m\AA\ ($\sigma_\lambda$) and km~s$^{-1}$ ($\sigma_v$). Note that the
  variances contain contributions from both statistical uncertainties and the inherent
  variability of the Sun. The difference between the literature wavelength and the new
  wavelength is given both in m\AA\ ($\delta_\lambda$) and km~s$^{-1}$
  ($\delta_\lambda$).}
\label{table:lambda}
\end{deluxetable}

  There are several difficulties with measuring Doppler velocities with EIS and it is
  important to outline the steps involved in computing them. One difficulty is the lack of
  absolutely calibrated rest wavelengths for the emission lines observed at these
  wavelengths. Tabulations of line identifications in the EIS wavelength ranges,
  \cite[e.g.,][]{brown2008}, use wavelengths from previous solar observations, such as the
  \cite{behring1976} rocket flight, and already include any systematic Doppler
  shifts. Thus even if the EIS wavelength calibration was known very precisely, velocity
  measurements relative to an absolute standard would still be impossible. The best we can
  do is to consider relative velocity measurements.

  There has been some work on the relative wavelengths of the emission lines measured with
  EIS \citep[e.g.,][]{brown2008}, but none that includes observations of the quiet corona
  above the limb, where we expect the line of sight velocities to be small for all
  lines. To investigate the relative wavelengths systematically we have analyzed data from
  another observing sequence that was run on the 18 May 2010 beginning at 11:14 UT. In
  this sequence a series of 120\,s exposures are taken with the 1\arcsec\ slit and the
  full CCD is telemetered to the ground. This series of deep exposures provide good
  statistics for all of the lines of interest. This means that the measured relative
  wavelengths are largely independent of the slit tilt and the orbital variation.  For
  each line of interest we have performed the same Gaussian fitting that was used for the
  slit raster data. For completeness we also consider several additional lines that were
  not included in the raster data. From these fits we calculate the median and the
  1-$\sigma$ variance in the measured centroid for each line. For each exposure the
  wavelengths are measured relative to the \ion{Fe}{12} 192.394\,\AA\ and \ion{Si}{10}
  258.375\,\AA\ lines.

  These new relative wavelengths are given in Table~\ref{table:lambda} and show that the
  centroids measured in the quiet corona above the limb are generally consistent with the
  values in the literature \citep{brown2008} to 4\,m\AA\ or better. There are, however,
  several important lines for which there are discrepancies. For example, the wavelength
  for the \ion{Fe}{13} 202.044\,\AA\ line needs to be adjusted to 202.0468\,\AA\ (a
  difference of 4.1\,km~s$^{-1}$) to provide velocities that are consistent with the
  \ion{Fe}{12} 192, 193, and 195\,\AA\ lines.  We also note that there is a typographical
  error in \cite{brown2008}, who give the wavelength for \ion{Si}{7} as
  275.352\,\AA. \cite{behring1976} give the wavelength of this line as 275.368\,\AA,
  consistent with our result of 275.3669\,\AA.

  Another correction that needs to be made accounts for the fact that the EIS slit is
  slightly tilted relative to the columns of the CCD. The magnitude of the slit tilt has
  been calculated for each pixel along the slit using extensive observations of emission
  lines above the limb in the quiet corona, where we expect the velocities to be uniform
  \citep{kamio2010}.

  Perhaps the most significant difficulty with EIS velocity measurements is that the
  spectra drift back and forth across the detector during the satellite orbit of 98
  minutes with an amplitude of around 2 wavelength pixels \citep[e.g.,][]{brown2007}. This
  behavior is thought to be due to the changing thermal environment of the satellite
  during an orbit. 

  To make an initial correction to the wavelength scale we use the spectral
  drift-spacecraft temperature model derived by \cite{kamio2010}. This model is based on
  the results from an artificial neural network that was trained on measurements of the
  \ion{Fe}{12} 195.119\,\AA\ centroid and temperature measurements taken within the
  instrument. This procedure corrects the wavelength scale based on the assumption that
  there are no net velocities in this line in the quiet corona. \cite{peter1999}, however,
  find that coronal lines are weakly blueshifted in the quiet Sun by around
  2--5\,km~s$^{-1}$. These values are much smaller than the high velocity upflows and
  downflows that we are concerned with in this paper.

  To further refine the wavelength scale we chose an emission line from each wavelength
  band and compute a spatially averaged line profile in a quiet region from each
  exposure. The wavelength scale is adjusted so there are no net velocities across the
  raster in these lines. For this procedure we use the \ion{Si}{7} 275.368\,\AA\ and
  \ion{Fe}{12} 195.119\,\AA\ lines. This procedure also verifies that the initial velocity
  correction has largely removed the temporal variability in the centroid. The differences
  between the centroid of the average profiles and the wavelength given in
  Table~\ref{table:lambda} are essentially constant with time. The systematic application
  of these new relative wavelengths and correction techniques to other EIS observations
  will be presented in a future paper.

  Given the extensive processing of the data needed to compute the line centroids, the
  absolute velocities should be considered with caution. The large downflows observed in
  \ion{Si}{7} and the large upflows observed \ion{Fe}{11}--\ion{Fe}{13} are very robust
  results. Some of the details are far less certain. The weak redshifts seen in
  \ion{Fe}{10}, for example, could easily be weak blueshifts. The conservative
  interpretation is that there is a transition between upflows and downflows in the
  temperature range where \ion{Fe}{9} and \ion{Fe}{10} are formed. The extension of the
  outflows to \ion{Fe}{15} should also be regarded with caution.

 \begin{figure*}[t!]
  \centerline{\includegraphics[clip,scale=1.0]{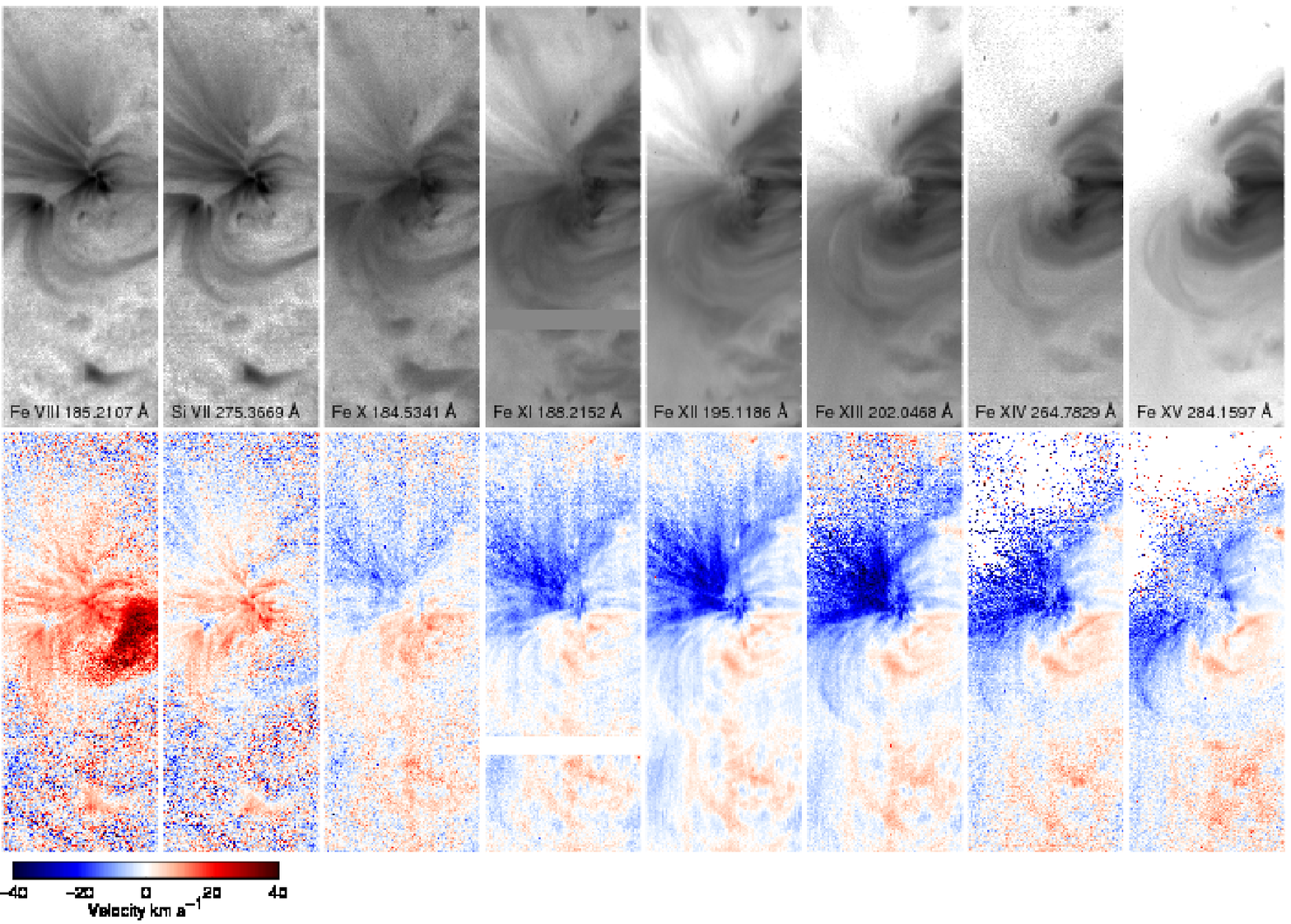}}
  \caption{EIS observations of active region outflows on 2007 February 20. This region has
    been considered by \protect{\cite{sakao2007}}, \protect{\cite{hara2008}},
    \protect{\cite{mcintosh2009}}, and \protect{\cite{he2010}}. The bright fan loops that
    are bright in \ion{Fe}{8} and \ion{Si}{7} are dominated by downflows while outflows
    are observed at higher temperatures. Note that \ion{Fe}{8} 185.213\,\AA\ is blended
    with a weak \ion{Ni}{16} line, which influences the fits in the high temperature core
    of the active region.}
  \label{fig:flows2}
  \end{figure*}

  \section{Summary and Discussion}

  We have presented systematic observations of an active region outflow observed with EIS
  using both narrow band slot imaging and slit rasters. These observations show that the
  outflow region has a complex velocity structure with strong downflows being observed at
  relatively cool temperatures and outflows being observed at higher temperatures. The
  presence of downflows on the fan loops suggests cooling plasma trapped on closed field
  lines. This interpretation is consistent with the larger field of view \textit{STEREO}
  EUVI images of this region.

  Earlier studies comparing outflows observed with EIS and simultaneous imaging data,
  \citep[e.g.,][]{sakao2007,hara2008,mcintosh2009,he2010}, have associated active region
  outflows with the apparent motions observed along the fan loops using broad-band imaging
  instruments, such as \textit{TRACE} 171\,\AA\ and 195\,\AA. Interestingly, these
  previous studies have all considered the same region but have not presented a full set
  of Doppler maps from EIS. In light of this we have analyzed data from this region using
  the methods outlined in the previous section. The intensity and Doppler maps, which are
  shown in Figure~\ref{fig:flows2}, are consistent with our results. The bright cool loops
  are dominated by inflows and the outflows occur at higher temperatures.
  
  So how do we form a coherent picture from these results? We conjecture that the fan
  loops and the outflows form two largely independent populations. In this view the fan
  loops imaged in \ion{Si}{7} are closed structures and some of the dynamics observed at
  higher temperatures are related to the heating and cooling of the plasma along these
  field lines (see \citealt{ugarte2009}). Furthermore, we speculate that most of the
  outflows lie on open field lines that connect to the heliosphere. The fact that the fan
  loops and the outflows tend to occur in the same general area of an active region could
  be related to the changes in magnetic topology that occur there
  \citep{baker2009,schrijver2010}.

%% ------------------------------------------------------------------------------------------
%% --- ACKNOWLEDGMENTS ----------------------------------------------------------------------
%% ------------------------------------------------------------------------------------------

\acknowledgments Hinode is a Japanese mission developed and launched by ISAS/JAXA, with
NAOJ as domestic partner and NASA and STFC (UK) as international partners. It is operated
by these agencies in co-operation with ESA and NSC (Norway).

%% ------------------------------------------------------------------------------------------
%% --- REFERENCES ---------------------------------------------------------------------------
%% ------------------------------------------------------------------------------------------

\end{document}